\documentclass[11pt]{article}
\usepackage{amssymb}

\begin{document}

\title{%
 $\kappa$--Rindler space}
\author{%
J. Kowalski-Glikman\\Institute for Theoretical Physics,\\
University of Wroclaw, Poland\footnote{e-mail:
jkowalskiglikman@ift.uni.wroc.pl} } \maketitle

\begin{abstract}
In this paper we construct, and investigate some thermal properties of, the non-commutative counterpart of Rindler space, which we call $\kappa$--Rindler space. This space is obtained by changing
variables in the defining commutators of $\kappa$--Minkowski space.
We then re-derive the commutator structure of $\kappa$--Rindler space with the help of an appropriate star product, obtained from the $\kappa$--Minkowski one. Using this star product, following the idea of Padmanabhan, we find the leading order, $1/\kappa$ correction to the Hawking thermal spectrum.
\end{abstract}


Classical general relativity is based on two principles. The first
is the equivalence principle which states that locally, in a
sufficiently small neighborhood of a given space-time point ${\cal
P}$  one can always describe physics in the language of special
relativity i.e., of particles and fields in Minkowski space. On the
basis of this principle one deduces that the dynamical field
describing gravity should be identified with a metric of space-time
manifold, or, more precisely, with the tetrad field (see e.g.,
\cite{Rovelli:2004tv} for detailed discussion.) The second postulate
of general relativity is the specific form of Einstein equations,
being a second order nonlinear differential equation for the metric,
or, equivalently, of Einstein--Hilbert action.

As it became more and more clear from investigations of classical
and  semi-classical properties of black holes there is one more
universal property of gravitating system: a deep relation between
horizons and thermodynamics. Already in early 1970s it was observed
by Bekenstein \cite{Bekenstein:1973ur} that four laws of black hole
dynamics, and especially the second one, can be given
thermodynamical interpretation, if one identifies entropy with the
area of the black hole horizon and temperature with its surface gravity
(acceleration of a static observer at the horizon, as seen by an
observer at infinity.) This observation has been but on firm ground
by  Hawking's seminal discovery of thermal black hole radiation
\cite{Hawking:1974sw}, with the temperature indeed proportional to
the surface gravity.

 It then turned out that there is a surprising relation between
thermodynamics of horizons and dynamics of gravity. In a remarkable
paper \cite{Jacobson:1995ab} (see also \cite{Padmanabhan:2002xm},
\cite{Padmanabhan:2003gd}) Jacobson was able to show that one can
interpret Einstein equations as  equations of state associated with
thermodynamics of a horizon that forms in flat space-time in the
coordinates associated with uniformly accelerating observer. Such
space-time is called Rindler space. It seems therefore that it is,
 Rindler, and not Minkowski, space that plays a
distinguished role in formulation of general relativity.

Rindler space can be obtained from Minkowski space with coordinates
$(T,X, Z_\bot)$, $Z_\bot=(Y,Z)$ by the following coordinates
transformation, corresponding to the observer moving with the
constant acceleration $a$
\begin{equation}\label{1}
    T=N(x)\sinh at,\quad X= N(x)\cosh at, \mathbf{Z}_\bot=\mathbf{z}_\bot
\end{equation}
where $N$ is a positive function of the coordinate $x$, under which
the Minkowski metric $ds^2=-dT^2+dX^2+dZ_\bot^2$  transforms to
\begin{equation}\label{2}
    ds^2 = -a^2 N^2(x)\, dt^2+(N')^2(x)dx^2 + d\mathbf{z}_\bot^2
\end{equation}
Clearly, the coordinates $(t,x, z_\bot)$ cover only part of the
original  Minkowski space, but one can extend them to cover the
whole of it. This reminds the situation one encounters analyzing
Kruskal coordinates for Schwarzschild black hole. To start with one
covers with these coordinates just the sector $I$ of black hole (the
space-time outside the horizon) and then one constructs maximally
extended coordinates covering all the remaining sectors. In fact the
analogy is even deeper, as one can easily check that Rindler space
describes the geometry of Schwarzschild black hole near the horizon,
if one identifies $a$ with the black hole surface gravity (see e.g.,
\cite{Padmanabhan:2003gd} and references therein.)

In spite of more than thirty years of intensive investigations
gravito-theormodynamics is still plagued with problems and puzzles,
of which the most well known ones are the mysterious origin of black
hole entropy and the information paradox. It is widely expected that we
will be able to handle them only when the theory of quantum gravity
is understood and Planck scale effects are properly taken into
account.

Although we still do not have quantum theory of gravity in our
disposal, there are reasons to believe that when quantum
gravitational effects are taken into account, space-time becomes
non-commutative, with the scale of non-commutativity of order of
Planck scale. This has been rather firmly established in the case of
2+1 gravity \cite{Freidel:2005bb} and there are indications that it
holds in the physical 3+1 gravity as well
\cite{KowalskiGlikman:2008fj}. Although not fully established, it is
claimed, by analogy with 2+1 dimensional result, that in 3+1
dimensions, the form of non-commutativity is that of
$\kappa$-Minkowski space \cite{Majid:1994cy},
\cite{Lukierski:1993wx} whose defining commutators in Cartesian
coordinates take the form
\begin{equation}\label{3}
    [\hat T, \hat X_i]=\frac{i}{\kappa}\, \hat X_i, \quad [\hat X_i, \hat X_j]=0
\end{equation}
where $\kappa$ is a mass scale, usually identified with Planck mass.

To find the accelerated observer counterpart of (\ref{3}) let us
make the  transformation analogous to (\ref{1}) (with $N(x)=x$)
\begin{equation}\label{4}
    \hat T=\hat x\, \sinh a \hat t, \quad \hat X=\hat x\, \cosh a \hat t, \quad \hat \mathbf{Z}_\bot=\hat \mathbf{z}_\bot
\end{equation}
By making use of the Baker--Campbell--Hausdorff formula one finds
the commutators of $(\hat t, \hat x)$ variables
\begin{equation}\label{5}
    [\hat t, \hat x]=\frac{i}{a\kappa}\, \cosh a \hat t,
\end{equation}
The commutators of these variables with the transversal ones are not easy  to find due to the notorious ordering problem. What one has to do is to deduce the commutators
$[\hat t, \hat \mathbf{z}_\bot]$ and $[\hat x, \hat \mathbf{z}_\bot]$ from the ones following from substituting (\ref{4}) to (\ref{3}), that is, from
\begin{equation}\label{6}
    [\hat x\, \sinh a \hat t, \hat \mathbf{z}_\bot]=\frac{i}\kappa\, \hat \mathbf{z}_\bot, \quad [\hat x\, \cosh a \hat t, \hat \mathbf{z}_\bot]=0,\quad [\hat \mathbf{z}_\bot, \hat \mathbf{z}_\bot]=0
\end{equation}
As we will see below (\ref{9b}), it can be expected that on the right hand side of the commutators $[\hat t, \hat \mathbf{z}_\bot]$ and $[\hat x, \hat \mathbf{z}_\bot]$ there will appear complex operators, involving $\hat x^{-1}$, which should be appropriately ordered. Unfortunately, we were not able to solve (\ref{6}) and find expressions for these commutators. Instead, in what follows we will make use of star product formalism, which is better suited to our present purposes.

As it is well known for $\kappa$-Minkowski non-commutative structure
there is an associated star product, whose construction is described
in details in \cite{Freidel:2007hk} (see also e.g.,
\cite{Kosinski:1999dw} and \cite{Daszkiewicz:2008bm}). The idea is
to replace a function on $\kappa$-Minkowski space $\hat f(\hat
\mathbf{X}, \hat T)$ with a corresponding function on ordinary
Minkowski space $f(\mathbf{X},T)$ with the help of the so-called
Weyl map ${\cal W}$
$$
{\cal W}\left[\hat f(\hat \mathbf{X}, \hat T)\right]=f(\mathbf{X},T)
$$
in such a way that the image under the Weyl map of a product of two
functions on $\kappa$-Minkowski space is given by star product of
their Minkowski space images, to wit
\begin{equation}\label{7}
  {\cal W}(\hat f(\hat \mathbf{X}, \hat T)\hat g(\hat \mathbf{X}, \hat T))=f(\mathbf{X},T)\star g(\mathbf{X},T)
\end{equation}
The $\star$ operator is rater complex, and its explicit form can be
found in \cite{KowalskiGlikman:2009zu}. For a moment we will need
just its leading order in the inverse powers of $\kappa$ expansion.
One finds
\begin{equation}\label{8}
    f\star g= fg -\frac{i}\kappa\left( T\frac{\partial f}{\partial \mathbf{X}}\cdot
    \frac{\partial g}{\partial \mathbf{X}}+\mathbf{X}\cdot
    \frac{\partial f}{\partial \mathbf{X}}\frac{\partial g}{\partial T}\right) + O(\kappa^{-2})
\end{equation}
where $\cdot$ denotes scalar product of three-vectors. To see how
this star product works let us reproduce the commutator (\ref{3}).
We  have
$$
[T,X]_\star\equiv T\star X_i - X_i\star T = TX_i - X_iT +\frac{i}\kappa
X_i=\frac{i}\kappa X_i
$$
as it should be. Since the higher order terms in $1/\kappa$ involve higher derivatives they do not contribute to the $[T,X]_\star$ commutator.

One can change variables in the star product (\ref{8}) to obtain
the one that is associated with $\kappa$-Rindler space. It reads
$$
f\star g(x, \mathbf{z}_\bot, t) = fg(x, \mathbf{z}_\bot, t) -
\frac{i}{a\kappa}\, \left(\cosh at \frac{\partial f}{\partial
x}-\frac{\sinh at}{ax} \frac{\partial f}{\partial t} \right)\,
\frac{\partial g}{\partial t} $$
\begin{equation}\label{9}
    -\frac{i}\kappa\, x \sinh at\, \frac{\partial f}{\partial \mathbf{z}_\bot}
    \cdot\frac{\partial g}{\partial \mathbf{z}_\bot}
    -\frac{i}\kappa\,\mathbf{z}_\bot\cdot \frac{\partial f}{\partial \mathbf{z}_\bot}
    \left(-\sinh at \frac{\partial g}{\partial x}+\frac{\cosh at}{ax} \frac{\partial g}{\partial t}\right)
\end{equation}

Using the star product we can define yet another commutator of basic variables. For example
\begin{equation}\label{9a}
    [t,x]_\star \equiv t\star x - x\star t =\frac{i}{a\kappa}\, \cosh at
\end{equation}
and thus the star commutator reproduces the formula we have found
above (\ref{5}).   Similarly we can find the star commutator
involving transversal coordinates $\mathbf{z}_\bot$. We have
\begin{equation}\label{9b}
    [t,\mathbf{z}_\bot]_\star \equiv t\star \mathbf{z}_\bot  - \mathbf{z}_\bot\star t =\frac{i}{ax\kappa}\, \mathbf{z}_\bot\, \cosh at
\end{equation}
\begin{equation}\label{9c}
    [x,\mathbf{z}_\bot]_\star \equiv x\star \mathbf{z}_\bot - \mathbf{z}_\bot\star x =-\frac{i}{a\kappa}\, \mathbf{z}_\bot\, \sinh at
\end{equation}
\begin{equation}\label{9d}
    [\mathbf{z}_\bot,\mathbf{z}_\bot]_\star  =0
\end{equation}
This provides us with the bracket structure for $\kappa$-Rindler space that we were not able to find in the operator formalism (see discussion above). This is because the star product  avoids the ordering problem, since the functions the operator (\ref{9}) acts on are already ordered, in the sense that they result from Weyl map of functions for whose some ordering has been chosen. We will discuss the relation between commutators of operators $(\hat x, \hat \mathbf{z}_\bot, \hat t)$ and star product commutators defined above in more details in the forthcoming paper. Notice however that the star product formalism, where we have to do with ordinary functions on space-time, whose physical meaning is clear, is much better suited for explicit model calculations than the operator formalism, where the meaning of $(\hat x, \hat \mathbf{z}_\bot, \hat t)$ and functions of them is far from clear.

Now we have all the necessary technical tools to address  the main
problem of this paper, namely the $\kappa$ corrections to some
gravito-thermodynamical effects.

Usually such question are addressed in the context of quantum field
theory in curved space-time. However, the core of the temperature
effects in gravitational fields (like Unruh effect or Hawking
radiation) is related to the fact that what inertial (or asymptotic
at ${\cal I}^-$) observer finds to be  a positive energy plane wave
becomes a thermal mixture of positive and negative frequencies when
observed by an accelerated one (asymptotic at ${\cal I}^+$). This
mixing of positive and negative frequencies can be, as noticed in
\cite{Padmanabhan:2003gd} (see also \cite{Alsing:2004ig}),
calculated in a purely classical context. Let us recall how this can
be done.

Consider the on-shell plane wave corresponding to a massless mode
with positive frequency $\Omega$ moving in $X$ direction in
Minkowski space
\begin{equation}\label{10}
    \phi(X,T)=\exp\left(-i\Omega(T-X)\right)
\end{equation}
Expressed in terms of the $(t,x)$ coordinates of accelerating
Rindler observer this plane wave has the form
\begin{equation}\label{11}
    \phi(X(x,t),T(x,t))\equiv\phi(x,t)=\exp\left(i\Omega x e^{-at}\right)
\end{equation}
which is clearly not monochromatic and instead has the frequency
spectrum $f(\omega)$, defined by Fourier transform
\begin{equation}\label{12}
    \phi(x,t)=\int_{-\infty}^{\infty}\frac{d\omega}{2\pi}\, f(\omega)\, e^{-i\omega t}
\end{equation}
with the associated power spectrum $P(\omega)=|f(\omega)|^2$. By
taking  the inverse Fourier transform we find
\begin{equation}\label{13}
    f(\omega)=\int_{-\infty}^{\infty} dt\, e^{i\Omega x e^{-at}}\, e^{i\omega t}
    =(\Omega x)^{i\omega/a}\, \Gamma\left(-\frac{i\omega}{a}\right)\, e^{\pi\omega/2a}
\end{equation}
Then one easily calculates that the power at negative frequencies,
per logarithmic band in frequencies is Planckian, at temperature
$T=a/2\pi$
\begin{equation}\label{14}
    \omega\, |f(-\omega)|^2=\frac{2\pi/a}{e^{2\pi\omega/a}-1}
\end{equation}
and the ratio of powers associated with negative and positive
frequency is
\begin{equation}\label{15}
   \frac{|f(-\omega)|^2}{|f(\omega)|^2}=e^{-2\pi\omega/a}
\end{equation}
Thus the  waves seen by the inertial observer as monochromatic with
arbitrary frequency  appears to the Rindler one as having thermal
spectrum. This observation is at the core of gravito-thermodynamics.

Our goal now is to find out what are the corrections to (\ref{14}),
(\ref{15}) resulting from noncommutative $\kappa$--Rindler space. To
set up the stage let us consider the integral in (\ref{13}).
Certainly we have to replace the ordinary product of functions
$e^{i\Omega x e^{-at}}\, e^{i\omega t}$ with the corresponding star
product
\begin{equation}\label{16}
    f^R_\kappa(\omega)=\int_{-\infty}^{\infty} dt\, e^{i\Omega(T(x,t)-X(x,t))}\star e^{i\omega t}
    =\int_{-\infty}^{\infty} dt\, e^{i\Omega x e^{-at}}\star e^{i\omega t}
\end{equation}
This formula is, of course,  not unambiguous. First it is not clear
why we choose the first term to be just the same plane wave as in
the undeformed case. The reason is that the star product has been
carefully chosen so that the Weyl image of $\kappa$-Minkowski
ordered plane wave, with time to the right $e^{-ik\hat X}e^{ik^0\hat
T}$ is just the standard plane wave (see \cite{Freidel:2007hk} for
detailed discussion)
$$
{\cal W}\left(e^{ik\hat X}e^{-ik^0\hat T}\right)=e^{-i(\Omega T - K X)}
$$
where $\Omega$, $K$ are functions  of the original labels $k, k^0$
satisfying for free massless field the condition $\Omega^2-K^2=0$,
which leads, after change of variables (\ref{1}) with $N(x)=x$ to
the expression in (\ref{16}). There is still the ambiguity of the
ordering of star product ($f\star g \neq g\star f$), but in this
paper we will consider that of (\ref{16}) only, because the star
product we are using was defined for the ``time to the right''
ordering.

Thanks to the fact that one of the terms in star product  (\ref{16})
is just a plane wave, corresponding to the massless mode, the
expression for it simplifies considerably. Using the explicit
formulas presented in \cite{KowalskiGlikman:2009zu} one
finds\footnote{As a result of different definition of what the
positive energy modes are, there is a sign difference between our
present convention and the one of \cite{KowalskiGlikman:2009zu}.}
\begin{equation}\label{17}
e^{i\Omega(T(x,t)-X(x,t))}\star e^{i\omega t}=    e^{i\Omega(T(x,t)-X(x,t))}
\lim_{{\cal T}, {\cal X}\rightarrow0} {\cal O} e^{i\omega t(T+{\cal T}, X+{\cal X})}
\end{equation}
with $t(T,X)\equiv 1/a\, \mbox{Ar}\tanh T/X$ and
\begin{equation}\label{18}
    {\cal O}\equiv \exp\left(-ixe^{-at}\Omega(E-1) -x\sinh at\,\frac{\Omega/\kappa}{1+\Omega/\kappa}\, F\right)
\end{equation}
where $E$ and $F$ are differential operators
\begin{equation}\label{19}
    E=-\frac{1}{i\kappa} \frac{\partial}{\partial{\cal T}}
    +\sqrt{-\frac{\Box}{\kappa^2}+1},\quad \Box = \frac{\partial^2}{\partial{\cal T}^2}- \frac{\partial^2}{\partial{\cal X}^2}
\end{equation}
\begin{equation}\label{20}
    F=\frac{\partial}{\partial{\cal T}}+ \frac{\partial}{\partial{\cal X}}
\end{equation}

These formulas are  the main result of the paper. It is rather clear
that there is little hope in finding explicit expressions from them
and below we will just calculate them in  the leading order in
$1/\kappa$ expansion. It is clear however from the form of
(\ref{17}), (\ref{18}) that contrary to the undeformed case
(\ref{14}) the power spectrum as seen by accelerated observer will
now depend on the original plane wave frequency $\Omega$ and the
observer distance from the origin $x$. This is easy to understand.
In the original calculation there was no scale and therefore all
values of $\Omega$ (and $x$) were `equally good.' In our case the
presence of $\kappa$ provides a reference point, with respect to
which $\Omega$ (and $x$) can be measured. For example it is now a
well posed question to ask if $\Omega$ is small as compared to the
Planck scale given by $\kappa$ or if the observer is at large
distance from the origin. Notice that the frequency (or energy)
dependence seems to be a general feature of $\kappa$ world whose
`rainbow' nature has been noticed in different contexts (see e.g.,
\cite{Magueijo:2002xx}, \cite{Arzano:2008bt}.)

After these general comments let  us return to calculating the
leading order corrections to the temperature spectrum. Expanding
${\cal O}$ we obtain
\begin{equation}\label{21}
    {\cal O} = 1 +\frac\Omega\kappa\left[ xe^{-at}\frac{\partial}{\partial{\cal T}}
    - x\sinh at\,\left(\frac{\partial}{\partial{\cal T}}+ \frac{\partial}{\partial{\cal X}}\right)\right]
    + O\left(\frac1{\kappa^2}\right)
\end{equation}
Acting with this operator on $e^{i\omega t}=\exp(i\omega/a\,
\mbox{Ar}\tanh T/X)$ as in (\ref{17}) and taking the limit we find
\begin{equation}\label{22}
    {\cal O}\, e^{i\omega t} = \left(1+\frac{i\omega\Omega}{a\kappa}\,
    e^{-2at}\right)e^{i\omega t} +O\left(\frac1{\kappa^2}\right)
\end{equation}
The counter part of the integral (\ref{13}) is now
$$
    f_\kappa(\omega)=\int_{-\infty}^{\infty} dt\, e^{i\Omega x e^{-at}}\, e^{i\omega t}\left(1+\frac{i\omega\Omega}{a\kappa}\,
    e^{-2at}\right)
$$
which can be easily evaluated with the help of the identity
$\Gamma(z+2)=z(z+1)\Gamma(z)$
\begin{equation}\label{23}
f_\kappa(\omega)=(\Omega x)^{i\omega/a}\,
\Gamma\left(-\frac{i\omega}{a}\right)\, e^{\pi\omega/2a} \left[1 -
\frac{i\omega}{a\kappa\Omega
x^2}\frac{i\omega}{a}\left(1-\frac{i\omega}{a}\right) \right]
\end{equation}
from this we deduce that the power, per logarithmic band in
frequency, at negative frequencies has the form
\begin{equation}\label{24}
    \omega\, |f_\kappa(-\omega)|^2=\frac1T\,
    \frac{1}{e^{\omega/T}-1}\left[1+\frac1\kappa\,
    \frac{T^2\omega^2}{2\pi^2\Omega x^2}\right]
\end{equation}
with $T\equiv a/2\pi$ is the Hawking temperature associated with the
acceleration. We see therefore that there are $1/\kappa$ no-thermal
corrections to the classical power spectrum. They form however differs from the ones calculated in the context of modified dispersion relation/generalized uncertainty principle in \cite{AmelinoCamelia:2005ik}, where the leading order corrections are appears only at $1/kappa^2$.
However, to this order
the ratio of powers of negative and positive frequencies is still
purely classical
\begin{equation}\label{25}
    \frac{|f_\kappa(-\omega)|^2}{|f_\kappa(\omega)|^2}=
    e^{-\omega/T}
\end{equation}
It is not clear if this last, quite surprising, result holds beyond the leading order approximation. This would be the case if ${\cal O} e^{i\omega t}$ would contain only even powers of $\omega$.

This question will be addressed in the forthcoming paper, where $\kappa$--Rindler space construction and its relation to boosts acting on $\kappa$--Minkowski space will be also presented.

\section*{Acknowledgment} I would like to thank Michele Arzano for his helpful comments
and for bringing \cite{Alsing:2004ig} to my attention. This research
was supported in parts
 by research projects N202 081 32/1844 and NN202318534 and
 Polish Ministry of Science and Higher Education grant 182/N-QGG/2008/0


\begin{thebibliography}{99}
\bibitem{Rovelli:2004tv}
  C.~Rovelli,
  ``Quantum Gravity,''
{\it  Cambridge, UK: Univ. Pr. (2004)}

\bibitem{Bekenstein:1973ur}
  J.~D.~Bekenstein,
  ``Black holes and entropy,''
  Phys.\ Rev.\  D {\bf 7} (1973) 2333.

\bibitem{Hawking:1974sw}
  S.~W.~Hawking,
  ``Particle Creation By Black Holes,''
  Commun.\ Math.\ Phys.\  {\bf 43}, 199 (1975)
  [Erratum-ibid.\  {\bf 46}, 206 (1976)].



\bibitem{Jacobson:1995ab}
  T.~Jacobson,
  ``Thermodynamics of space-time: The Einstein equation of state,''
  Phys.\ Rev.\ Lett.\  {\bf 75} (1995) 1260
  [arXiv:gr-qc/9504004].

\bibitem{Padmanabhan:2002xm}
  T.~Padmanabhan,
  ``Is gravity an intrinsically quantum phenomenon? Dynamics of gravity  from
  the entropy of spacetime and the principle of equivalence,''
  Mod.\ Phys.\ Lett.\  A {\bf 17}, 1147 (2002)
  [arXiv:hep-th/0205278].

\bibitem{Padmanabhan:2003gd}
  T.~Padmanabhan,
  ``Gravity and the thermodynamics of horizons,''
  Phys.\ Rept.\  {\bf 406}, 49 (2005)
  [arXiv:gr-qc/0311036].

\bibitem{Freidel:2005bb}
  L.~Freidel and E.~R.~Livine,
  ``Ponzano-Regge model revisited. III: Feynman diagrams and effective  field
  theory,''
  Class.\ Quant.\ Grav.\  {\bf 23} (2006) 2021
  [arXiv:hep-th/0502106].
  L.~Freidel and E.~R.~Livine,
 ``Effective 3d quantum gravity and non-commutative quantum field theory,''
  Phys.\ Rev.\ Lett.\  {\bf 96}, 221301 (2006)
  [arXiv:hep-th/0512113].

\bibitem{KowalskiGlikman:2008fj}
  J.~Kowalski-Glikman and A.~Starodubtsev,
  ``Effective particle kinematics from Quantum Gravity,''
  Phys.\ Rev.\  D {\bf 78} (2008) 084039
  [arXiv:0808.2613 [gr-qc]].

  \bibitem{Majid:1994cy}
  S.~Majid and H.~Ruegg,
  ``Bicrossproduct Structure Of Kappa Poincare Group And Noncommutative
  Geometry,''
  Phys.\ Lett.\  B {\bf 334} (1994) 348
  [arXiv:hep-th/9405107].


\bibitem{Lukierski:1993wx}
  J.~Lukierski, H.~Ruegg and W.~J.~Zakrzewski,
  ``Classical Quantum Mechanics Of Free Kappa Relativistic Systems,''
  Annals Phys.\  {\bf 243} (1995) 90
  [arXiv:hep-th/9312153].

  \bibitem{Freidel:2007hk}
  L.~Freidel, J.~Kowalski-Glikman and S.~Nowak,
  ``Field theory on $\kappa$--Minkowski space revisited: Noether charges and
  breaking of Lorentz symmetry,''
  Int.\ J.\ Mod.\ Phys.\  A {\bf 23} (2008) 2687
  [arXiv:0706.3658 [hep-th]].

\bibitem{Kosinski:1999dw}
  P.~Kosinski, J.~Lukierski and P.~Maslanka,
  ``Local field theory on kappa-Minkowski space, star products and
  noncommutative translations,''
  Czech.\ J.\ Phys.\  {\bf 50}, 1283 (2000)
  [arXiv:hep-th/0009120].


\bibitem{Daszkiewicz:2008bm}
  M.~Daszkiewicz, J.~Lukierski and M.~Woronowicz,
  ``Kappa-deformed oscillators, the choice of star product and free
  kappa-deformed quantum fields,''
  arXiv:0807.1992 [hep-th].

\bibitem{KowalskiGlikman:2009zu}
  J.~Kowalski-Glikman and A.~Walkus,
  ``Star product and interacting fields on $\kappa$-Minkowski space,''
  arXiv:0904.4036 [hep-th].

\bibitem{Alsing:2004ig}
  P.~M.~Alsing and P.~W.~Milonni,
  ``Simplified derivation of the Hawking-Unruh temperature for an  accelerated
  observer in vacuum,''
  Am.\ J.\ Phys.\  {\bf 72} (2004) 1524
  [arXiv:quant-ph/0401170].



\bibitem{Magueijo:2002xx}
  J.~Magueijo and L.~Smolin,
  ``Gravity's Rainbow,''
  Class.\ Quant.\ Grav.\  {\bf 21} (2004) 1725
  [arXiv:gr-qc/0305055].


\bibitem{Arzano:2008bt}
  M.~Arzano and D.~Benedetti,
  ``Rainbow statistics,''
  arXiv:0809.0889 [hep-th].
  
\bibitem{AmelinoCamelia:2005ik}
  G.~Amelino-Camelia, M.~Arzano, Y.~Ling and G.~Mandanici,
  ``Black-hole thermodynamics with modified dispersion relations and
  generalized uncertainty principles,''
  Class.\ Quant.\ Grav.\  {\bf 23} (2006) 2585
  [arXiv:gr-qc/0506110].




\end{thebibliography}
\end{document}